\documentclass[letterpaper, 10 pt, conference]{ieeeconf} 
\IEEEoverridecommandlockouts
\usepackage{graphics} 
\usepackage{graphicx}   
\graphicspath{{Legacy/}}
\usepackage{amsmath} 

\usepackage[font={footnotesize}]{caption}
\usepackage{mathtools, cuted}
\usepackage{interval}
\usepackage{nicefrac}
\usepackage[noadjust]{cite}

\usepackage{booktabs}

\usepackage{tikz}
\usetikzlibrary{positioning}

\usepackage{etoolbox}
\makeatletter
\patchcmd{\@makecaption}
  {\scshape}
  {}
  {}
  {}
\makeatletter
\patchcmd{\@makecaption}
  {\\}
  {.\ }
  {}
  {}
\makeatother

\usepackage{siunitx}
\usepackage{breqn}
\usepackage{lscape}
\usepackage{derivative}
\usepackage{nicematrix}
\usepackage{color, colortbl}
\usepackage{multirow}
\usepackage{nicefrac}

\usepackage[english]{babel}
\usepackage{amssymb}  

\usepackage{hyperref}

\renewcommand{\vec}[1]{\mathbf{\boldsymbol{#1}}}

\title{\LARGE \bf Towards Dynamic Quadrupedal Gaits: A Symmetry-Guided RL \\ Hierarchy Enables Free Gait Transitions at Varying Speeds}

\author{Jiayu Ding$^*$, Xulin Chen$^*$, Garrett E. Katz, and Zhenyu Gan
\thanks{Jiayu Ding and Zhenyu Gan are with the Department of Mechanical and Aerospace Engineering, Syracuse University, Syracuse, NY 13244 \texttt{\{ jding14, zgan02\}@syr.edu}.
Xulin Chen and Garrett E. Katz are with the Department of Electrical Engineering and Computer Science, Syracuse University \texttt{\{ xchen168, gkatz01\}@syr.edu}.
Jiayu Ding$^*$ and Xulin Chen$^*$ contributed equally to this publication.}
\thanks{This work was supported by a startup fund from the Syracuse University.}
 }

\begin{document}

\maketitle
\thispagestyle{plain}
\pagestyle{plain}

\begin{abstract}
Quadrupedal robots exhibit a wide range of viable gaits, but generating specific footfall sequences often requires laborious expert tuning of numerous variables, such as touch-down and lift-off events and holonomic constraints for each leg. This paper presents a unified reinforcement learning framework for generating versatile quadrupedal gaits by leveraging the intrinsic symmetries and velocity-period relationship of dynamic legged systems. We propose a symmetry-guided reward function design that incorporates temporal, morphological, and time-reversal symmetries. By focusing on preserved symmetries and natural dynamics, our approach eliminates the need for predefined trajectories, enabling smooth transitions between diverse locomotion patterns such as trotting, bounding, half-bounding, and galloping. Implemented on the Unitree Go2 robot, our method demonstrates robust performance across a range of speeds in both simulations and hardware tests, significantly improving gait adaptability without extensive reward tuning or explicit foot placement control. This work provides insights into dynamic locomotion strategies and underscores the crucial role of symmetries in robotic gait design.

\end{abstract}


\section{Introduction}
\label{sec: Intro}

Quadrupedal robots hold promise for applications such as search-and-rescue, industrial inspection, and planetary exploration. A fundamental limitation, however, is the absence of a unified framework that enables these robots to generate and switch among diverse gaits on demand. In animals, such transitions are routine: walking conserves energy, running increases speed, and galloping clears obstacles. Replicating this adaptive capability in robots would greatly expand their effectiveness in unstructured environments. Yet despite decades of research, most robotic systems remain restricted to a small set of pre-programmed gaits. Developing controllers that can flexibly coordinate leg motions across gait classes therefore remains a central challenge in legged locomotion research.  
Most existing approaches fall short of this goal. Trajectory optimization and model predictive control ~\cite{2019minicheetahMPC} rely on fixed, hand-coded footfall sequences that work well in structured conditions but degrade under uncertainty. Reinforcement learning (RL) has emerged as a powerful alternative, but current methods typically require massive hand-tuning for reward function design \cite{fu2023deep}. Central pattern generator (CPG)-based controllers produce multiple gaits, but they depend on explicitly prescribed foot trajectories that may not match the robot’s intrinsic dynamics. 


Symmetry offers a solution to harness the advantage of both approaches. Hildebrand’s taxonomy~\cite{Hildebrand1977} classified quadrupedal gaits in terms of temporal and spatial symmetries, while later robotics studies~\cite{razavi2017symmetry,Raibert1986symmetry} showed that gaits can be derived from compositions and disruptions of temporal, spatial, and morphological symmetries. This perspective treats gaits not as isolated behaviors but as related members of broader families, and embedding symmetry into learning provides a compact representation of gait repertoires while reducing the search space for policy training.

\begin{figure}[t]
\centering
\includegraphics[width=0.9\columnwidth]{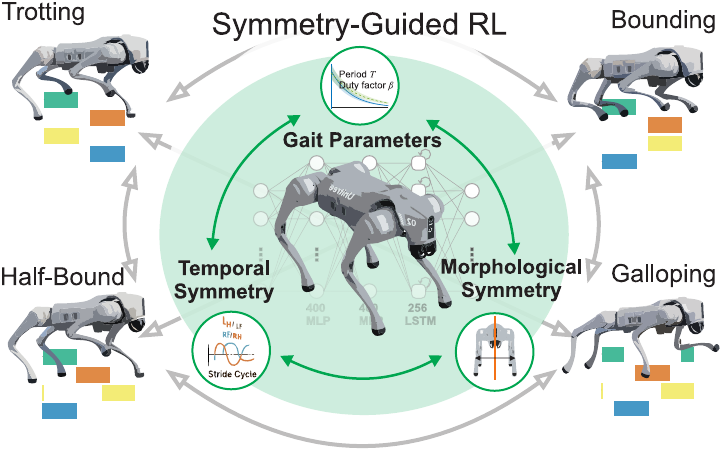}
\caption{Concept overview of the symmetry-guided reinforcement learning framework. User commands and gait parameters feed an MLP policy with temporal and morphological symmetries, enabling a single policy to generate trotting, bounding, half-bounding, and galloping on the Unitree Go2 without predefined trajectories.}
\label{fig:hardware_gaits}
\vspace*{-0.2in}
\end{figure}

\begin{figure*}[t]
\centering
\vspace*{-0.1in}
\includegraphics[width=1\textwidth]{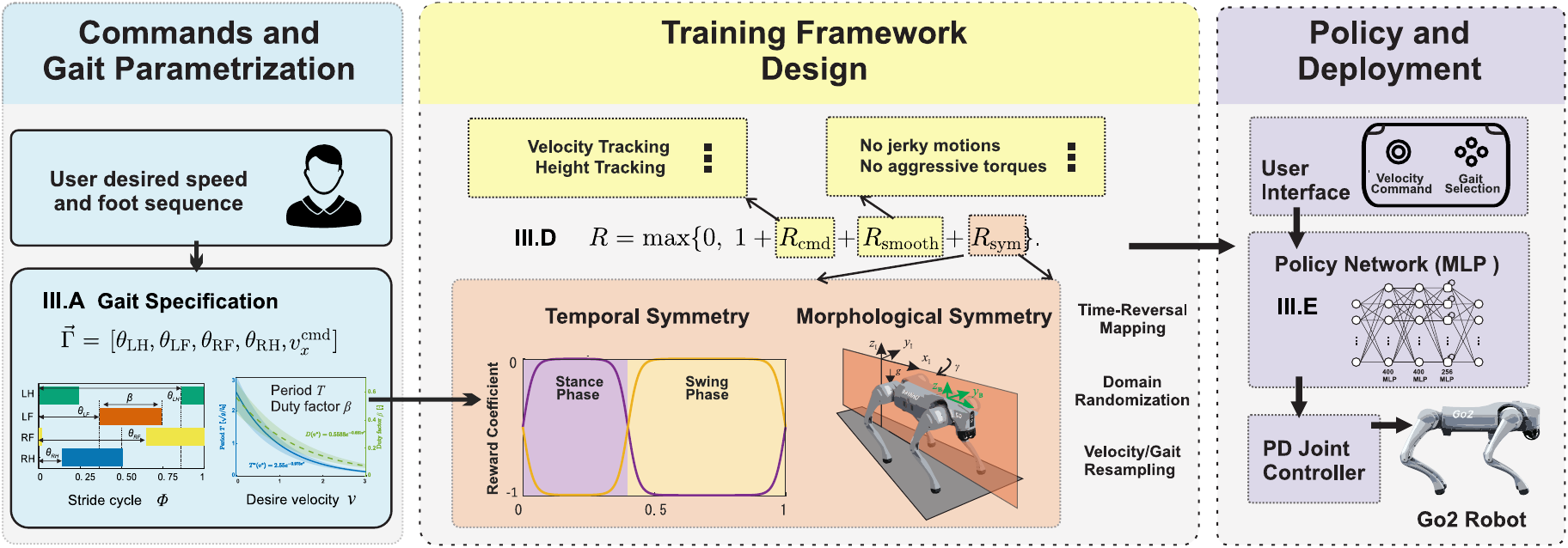}
\caption{Overview of the proposed framework. \textbf{Left:} User commands ($v_x^{\text{cmd}}, v_y^{\text{cmd}}, \omega_{\text{yaw}}^{\text{cmd}}$) and selected gait sequences (trotting, bounding, half-bounding, galloping) are mapped into gait parameters $\vec{\Gamma} = [\theta_{\text{LH}}, \theta_{\text{LF}}, \theta_{\text{RF}},  \theta_{\text{RH}}, v_x^{\text{cmd}}]$. 
\textbf{Middle:} Training framework design integrates reward function design (command tracking, smoothness, temporal and morphological symmetry), along with time-reversal mapping, domain randomization, and velocity/gait resampling. 
\textbf{Right:} The framework drives training of our MLP policy network, which outputs joint targets tracked by a PD controller on the Unitree Go2 robot, with a user interface providing real-time command input.} 
\vspace*{-0.2in}
\label{fig:training_structure}
\end{figure*}
%

In this paper, we present a symmetry-guided reinforcement learning framework for quadrupedal gait generation. Temporal, morphological, and time-reversal symmetries are incorporated directly into the reward function and phase mappings, guiding policy learning without predefined sequences or gait-specific tuning. Using this approach, we train a single policy that reproduces trotting, bounding, half-bounding, and galloping on the Unitree Go2 robot (Fig.~\ref{fig:hardware_gaits}). We evaluate the learned policy in both simulation and hardware across a wide range of commanded speeds and gait transitions, using metrics for velocity tracking accuracy, gait consistency, and energy efficiency. Results show that symmetry-enforced policies achieve more accurate velocity regulation, more coordinated footfall patterns, and reduced cost of transport compared to baselines without symmetry.

The contributions of this work are:
1) A symmetry-constrained reward design that simplifies gait learning by reducing the dimensionality of command space. 
2) A unified policy that reproduces a spectrum of quadrupedal gaits on hardware, demonstrating improved tracking, consistency, and efficiency over policies trained without symmetry.
By incorporating symmetries into reinforcement learning, this work establishes a scalable framework for adaptive quadrupedal locomotion, offering new insights into dynamic gait generation and control.

\section{Related Works}
\label{sec:related works}

\subsection{Symmetry in Locomotion and Control}
Symmetry has long served as a foundation for locomotion analysis and controller design. Hildebrand \cite{Hildebrand1965,Hilderbrand1989Quadrupedal} classified quadrupedal gaits by inter-limb phase relations, showing that symmetrical and asymmetrical gaits form continuous families that enable smooth transitions. Raibert \cite{Raibert1986symmetry} emphasized time-reversal symmetry in running robots, where forward and backward motions are mirror images. Razavi \emph{et al.} \cite{razavi2017symmetry} applied odd–even symmetry to generate efficient periodic gaits, and Ordonez \emph{et al.} \cite{ordonez2023discrete} demonstrated that exploiting discrete symmetries improves neural-network sample efficiency. Ding \emph{et al.} \cite{Ding2024} showed that symmetry breaking expands quadrupedal gait diversity. More recently, Su \emph{et al.} \cite{su2024leveraging} incorporated equivariance constraints into RL architectures, improving gait quality and sim-to-real robustness. Symmetry has also been exploited in perception tasks, where Butterfield \emph{et al.} \cite{butterfield2025mihgnn} designed a morphology-informed graph neural network for contact estimation, highlighting its broader role in robot learning.

\subsection{Reinforcement Learning for Locomotion}
RL-based locomotion methods can be categorized into reference-based and reference-free approaches. Reference-based methods rely on predefined templates. On Cassie, a SLIP-based gait library supported walking up to 1.2 m/s \cite{Green2021SLIPGuideRL}, and RL feedback enabled robust disturbance recovery \cite{Xie2018CassieWalkingRL}. In quadrupeds, CPG frameworks encode phase shifts and duty factors to parameterize cyclic motions \cite{Schner1990,shao2021learning}, but they constrain robots to handcrafted templates. Reference-free methods instead rely on enforcing specific properties of gaits. Early works were limited to single gaits, such as walking towards a certain direction \cite{Haarnoja2019LearningtoWalk} or trotting based on specified commands \cite{Kohl2004Aibostrot}. Later, Siekmann \emph{et al.} \cite{Siekmann2021CassieAllBipedalGaits} enforced the periodicity of foot ground reaction forces and velocities to learn all bipedal gaits, and Margolis and Agrawal \cite{margolis2023walk} expanded the command space to diversify quadrupedal gaits. These designs achieve variety but require careful reward tuning to maintain training stability.

Although symmetry has proven effective for both locomotion analysis and learning, most RL-based gait generation still depends on explicit models or command-space augmentation. Our framework differs by embedding temporal, time-reversal, and morphological symmetries directly into the reward design. This inductive bias enforces structural invariances of gait families, reduces the number of gait parameters, and enables dynamic quadrupedal gaits that generalize across speeds and transitions without predefined foot sequences.

\section{Symmetry-Guided Reinforcement Learning}
\label{methods}

We present a unified framework to explore quadrupedal locomotion across symmetry classes using a set of principal gait parameters. The method integrates symmetry-aware gait specification, a Markov decision process formulation, structured state–action design, and a reward combining command tracking, smoothness, and symmetry terms. As illustrated in Fig.~\ref{fig:training_structure}, this yields robust policies that generalize across gaits and commanded speeds.

\subsection{Gait Specification}
\label{subsec:gait_spec}

\begin{figure}[bht]
\centering
\vspace*{-0.10in}
\includegraphics[width=1\columnwidth]{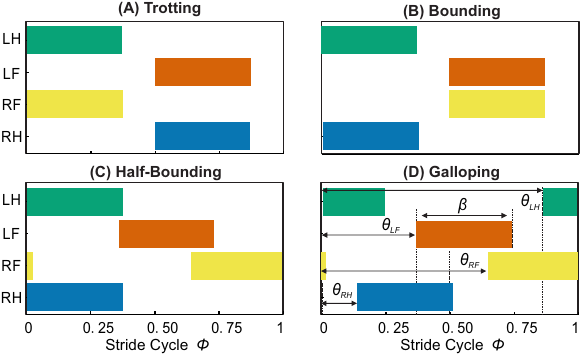}
\caption{Footfall patterns of four quadrupedal gaits. Colored bars denote stance and blanks denote swing. {\footnotesize LH, LF, RF, RH} indicate left hind, left front, right front, and right hind legs. These examples span the symmetry classes analyzed.}
\vspace*{-0.0in}
\label{fig:gait_examples}
\end{figure}

A systematic parameterization of quadrupedal gaits provides the foundation for generating trajectories and analyzing symmetry properties in our framework. We model a gait as the periodic orbit of a hybrid dynamical system~\cite{Ding2024}, where each cycle alternates between stance and swing phases over a stride \emph{period} $T$. The normalized \emph{phase} $\phi \in [0,1)$ denotes the position within the cycle, and the phase of each leg $i \in \{\text{LH}, \text{LF}, \text{RF}, \text{RH}\}$ (left hind, left front, right front, right hind, respectively) is shifted by a leg-specific \emph{phase offset} $\theta_i \in [0,1)$. The fraction of the stride in stance is described by the \emph{duty factor} $\beta \in (0,1)$. 
The stride period $T$ and duty factor $\beta$ are intrinsically coupled to forward velocity. Fixing them limits attainable speeds, while arbitrary randomization hinders convergence. Prior studies~\cite{GanDynamicSimilarity, Ding2024, alqaham2023energetic} showed that these temporal parameters align with oscillatory modes from body–limb energy exchanges. To capture this dependency, we combine insights from the passive dynamics of a quadrupedal SLIP model~\cite{gan2016passive, Ding2024} with empirical calibration from RL trials, yielding
\begin{align}
    T^* &= a_T (1 + b_T \delta |v_x^{*\ \text{cmd}}|)\, e^{-c_T |v_x^{*\ \text{cmd}}|}, \nonumber \\
    \beta &= a_\beta (1 + b_\beta \delta |v_x^{*\ \text{cmd}}|)\, e^{-c_\beta |v_x^{*\ \text{cmd}}|},
    \label{eq:gait-period-and-duty-factor-expression}
\end{align}
where $\delta \sim  U(-1,1)$ introduces small uniform perturbations, $v_x^{*\ \text{cmd}}$ is the command velocity normalized by $\sqrt{gl}$, and $g$ and $l$ are gravity and leg length, respectively. The fitted constants $(a_T, b_T, c_T, a_\beta, b_\beta, c_\beta)$ are reported in Section \ref{subsec:RLsetup}. For tractability, we assume $\beta$ is identical across legs. We analyze four gait families with symmetry-based variants (Fig.~\ref{fig:gait_examples}), summarized in Table~\ref{tab:gait_phases}.
These symmetry distinctions determine the structure of feasible trajectories and serve as constraints in subsequent optimization. Under the above definitions and assumptions, we specify a gait using its leg phase offsets and forward command velocity
\begin{equation}
    \vec{\Gamma} \coloneq \left[ \theta_{\text{LH}}, \theta_{\text{LF}}, \theta_{\text{RF}}, \theta_{\text{RH}}, v_x^{\text{cmd}} \right].
\end{equation}

\begin{table}[t]
    \centering
    \renewcommand{\arraystretch}{1.2}
    \begin{tabular}{lcccc}
        \toprule
        \textbf{Gait} & $\theta_{\text{LH}}$ & $\theta_{\text{LF}}$ & $\theta_{\text{RF}}$ & $\theta_{\text{RH}}$ \\
        \midrule
        Trotting       & 0.00 & 0.50 & 0.00 & 0.50 \\
        Bounding       & 0.00 & 0.50 & 0.50 & 0.00 \\
        Half-bounding (L) & 0.00 & 0.63 & 0.37 & 0.00 \\
        Half-bounding (R) & 0.00 & 0.37 & 0.63 & 0.00 \\
        Rotary gallop  & 0.13 & 0.37 & 0.63 & 0.87 \\
        Transverse gallop & 0.13 & 0.63 & 0.37 & 0.87 \\
        \bottomrule
    \end{tabular}
    \caption{Phase offsets $\theta_i$ for representative quadrupedal gaits. LH, LF, RF, RH denote left hind, left front, right front, and right hind legs.}
    \label{tab:gait_phases}
    \vspace*{-0.20in}
\end{table}

\subsection{Problem Formulation}
We formulate our problem within the context of a discrete-time Markov Decision Process (MDP) defined by the tuple $(\mathcal{S}, \mathcal{A}, R, P, \gamma)$. In an MDP, $\mathcal{S}$ and $\mathcal{A}$ represent the continuous state and action space respectively, and $\gamma \in (0, 1)$ denotes the discount factor. At any time step $t$, the agent selects an action $\vec{a}_t \in \mathcal{A}$ to move from the current state $\vec{s}_t \in \mathcal{S}$ to the next state $\vec{s}_{t+1} \in \mathcal{S}$, following the transition probability $P(\vec{s}_{t+1}|\vec{s}_t,\vec{a}_t)$ and returning a scalar reward $R(\vec{s}_t,\vec{a}_t)$. The goal is to discover the optimal control policy $\pi(\vec{a}_t|\vec{s}_t)$ that maximizes the expectation of the discounted return $\mathcal{J}(\pi)$ over any trajectories induced by $\pi$:
\begin{equation}
    \mathcal{J}(\pi) = \mathbb{E}_{\vec{a}_t\sim \pi(\cdot|\vec{s}_t)} \left[ \sum_{t=0}^\infty \gamma^t R(\vec{s}_t,\vec{a}_t) \right].
    \label{eq:obj-of-mdp}
\end{equation}

\subsection{State and Action Space}
We implement our framework on a 12 degree-of-freedom Unitree quadrupedal robot Go2, where each leg has a hip, thigh and calf joint. The action space $\mathcal{A} \in \mathbb{R}^{12}$ is the target position of 12 joints, and the position control is achieved via a PD controller with $k_p=30$ and $k_d=0.65$. The state space $\mathcal{S} \in \mathbb{R}^{52}$ encompasses the observed robot states, gait parameters and user commands. To be specific, the state vector $\vec{s}_t \in \mathcal{S}$ at time step $t$ includes proprioception (orientation $\vec{g}^{\text{ori}} \in \mathbb{R}^3$, joint positions $\vec{q} \in \mathbb{R}^{12}$ and velocities $\dot{\vec{q}} \in \mathbb{R}^{12}$), last action $\vec{a}_{t-1}  \in \mathbb{R}^{12}$, velocity commands $(v^{\text{cmd}}_x,v^{\text{cmd}}_y,\omega^{\text{cmd}}_{\text{yaw}})$, the clock input of four legs $\vec{p}=\{\sin (2\pi\phi_i) ), i \in \{ \text{LH, LF, RF, RH} \} \}$ using the current leg phase $\phi_i$, the phase offset of legs for indicating the desired gait $\vec{\theta} = [\theta_{\text{LF}},\theta_{\text{RF}},\theta_{\text{LH}},\theta_{\text{RH}}]$, and the ratio of stance and swing phase $\vec{r}=[\beta, 1-\beta]$.

\subsection{Reward Function Design}
\label{subsec:rewardsesign}

As shown in Fig.~\ref{fig:training_structure}, the reward function combines multiple objectives: tracking user commands, encouraging smooth and energetically efficient motions, and enforcing symmetries. Each component is scaled to similar magnitudes to stabilize training, and the total reward is clipped to be non-negative:
\begin{equation}
R = \max\{0,\; 1 + R_{\text{cmd}} + R_{\text{smooth}} + R_{\text{sym}}\}.
\end{equation}

\begin{table}[t]
    \centering
    \renewcommand{\arraystretch}{1.2}
    \begin{tabular}{l l}
        \toprule
        \textbf{Term} & \textbf{Expression} \\
        \midrule
        $x$ velocity tracking & $-0.3\big(1-\exp\{-2|v_x-v_x^{\text{cmd}}|\}\big)$ \\
        $y$ velocity tracking & $-0.3\big(1-\exp\{-10|v_y-v_y^{\text{cmd}}|\}\big)$ \\
        yaw velocity tracking & $-0.3\big(1-\exp\{-5|\omega_{\text{yaw}}-\omega_{\text{yaw}}^{\text{cmd}}|\}\big)$ \\
        base height tracking & $-0.3\big(1-\exp\{-5d_h\}\big)$ \\
        \midrule
        torque differences & $-0.1\big(1-\exp\{-0.1\|\vec{\tau}_t-\vec{\tau}_{t-1}\|_1\}\big)$ \\
        hip action smoothness & $-0.15\big(1-\exp\{-0.5\sum |\vec{a}_t^{\text{hip}}|\}\big)$ \\
        swing-phase clearance & $-0.15I_{\text{swing}}(\phi_i)\big(1-\exp\{-20c_{\text{foot}}\}\big)$ \\
        \bottomrule
    \end{tabular}
    \caption{Reward terms for command tracking $R_{\text{cmd}}$ (top) and smoothness $R_{\text{smooth}}$ (bottom).}
    \label{tab:cmd-smoothness-reward}
    \vspace*{-0.2in}
\end{table}

Table~\ref{tab:cmd-smoothness-reward} introduces all the terms included in the command tracking reward $R_{\text{cmd}}$ and the smoothness reward $R_{\text{smooth}}$. 
The command tracking reward $R_{\text{cmd}}$ penalizes deviations from commanded base velocities in $x$, $y$, and yaw, and encourages the torso height $h_{\text{base}}$ to remain within the range 
$[h^{\min}_{\text{base}},\,h^{\max}_{\text{base}}]$ by penalizing boundary violations:
\begin{equation}
d_h = \max\{0,\, h^{\min}_{\text{base}} - h_{\text{base}}\}
     + \max\{0,\, h_{\text{base}} - h^{\max}_{\text{base}}\}.
\end{equation}

The smoothness reward $R_{\text{smooth}}$ penalizes rapid torque changes, excessive hip motions, and insufficient foot clearance during swing. 
Here $\vec{a}_t^{\text{hip}}$ is the softmax-weighted magnitude of hip joint actions, the binary indicator $I_{\text{swing}}(\phi_i)$ equals $1$ if leg $i$ is in swing at phase $\phi_i$ and $0$ otherwise, and $c_{\text{foot}}$ quantifies clearance shortfall:
\begin{align}
c_{\text{foot}} = \sum_{i=1}^{4} 
w(s_i)
\max\bigl(0,\,h^{\min}_{\text{cl}} - z_i \bigr), \\
w(s_i) \;=\; \tfrac{1}{2}\Big(1 + \sin(\pi s_i)\Big) \notag   
\end{align}
where $z_i$ is the vertical position of foot $i$, $h^{\min}_{\text{cl}}$ the desired minimal clearance, and $\ s_i \;=\; \operatorname{clip}\!\left(\frac{\phi_i}{\,1-\beta\,},\,0,\,1\right)$.

Besides, we incorporate three types of symmetry observed in animal and robotic locomotion: temporal, morphological, and time-reversal. Time-reversal symmetry is not expressed as an explicit reward but enforced via phase mapping, and the final symmetry reward is defined as $R_{\text{sym}} = R_{\text{tem}} + R_{\text{mor}}$.

\subsubsection{Temporal symmetry}
To ensure each leg exhibits a single stance per stride and avoid Zeno-like switching~\cite{ames2005sufficient}, we gate penalties with phase-dependent stance/swing indicators. For leg $i$, stance is defined over $[1-\beta,1)$ and swing over $[0,1-\beta)$. To smooth the phase transition, we introduce the Von Mises distribution to the binary indicators $I_{\text{swing}}$ and $I_{\text{stance}}$ and calculate their expectation \cite{Siekmann2021CassieAllBipedalGaits}. The reward penalizes high velocity when feet are desired to be stance, and nonzero ground reaction forces (GRFs) when feet are desired to be swing
\begin{equation}
\begin{aligned}
R_{\text{tem}}=-0.15 & \sum_i \Big(
\mathbb{E}\!\left[I_{\text{swing}}(\phi_i)\right]\!\bigl(1-\exp\{-0.001\lVert\vec f_i\rVert\}\bigr)\\
&+\,\mathbb{E}\!\left[I_{\text{stance}}(\phi_i)\right]\!\bigl(1-\exp\{-2\lVert\vec v_i\rVert\}\bigr)
\Big),
\end{aligned}
\end{equation}
with $\vec{f}_i$ the GRF and $\vec{v}_i$ the foot velocity of leg $i$.

\subsubsection{Morphological symmetry}
When two legs share the same phase offset ($\theta_i=\theta_j$), they should produce similar joint trajectories. We use a tolerance $\epsilon_\sigma=0.01$ to detect such symmetry and penalize deviations:
\begin{equation}
\begin{aligned}
R_{\text{mor}} &= -0.15 \Big(1 - \exp\{-5 d(G_\sigma)\}\Big), \\
d(G_\sigma) &= \sum_{\substack{\sigma(i,j)\in G_\sigma \\ k \in \{\text{hip}, \text{thigh}, \text{knee}\}}} 
   f(\sigma(i,j)) \, | q_{i,k} - q_{j,k} |,
\end{aligned}
\end{equation}
where $q_{(\cdot,\cdot)}$ are joint positions and $f(\sigma(i,j))=1$ if $|\theta_i-\theta_j|\leq\epsilon_\sigma$ and 0 otherwise. This discourages uneven leg usage and prevents limping behaviors.

\subsubsection{Time-reversal symmetry}
Quadrupedal locomotion exhibits approximate invariance under reversing time and direction~\cite{Raibert1986symmetry, Ding2024}. To enforce this property for leg $i$, we remap its phase in time-reversal style when the commanded velocity is negative, and the modified leg phase $\phi_i$ is 
\begin{equation}
\label{eq:time-reversal-map}
\phi_i = 
    \left\{ \begin{array}{lcl}
(\phi + \theta_i) \ \mbox{mod} \ 1 & \mbox{if}
& v^{\text{cmd}}_x \geq 0 \\ 
-(\phi + \theta_i) \ \mbox{mod} \ (-1) & \mbox{if} & v^{\text{cmd}}_x < 0
\end{array}\right.
\end{equation}
where $\mbox{mod}$ is the modulo function. This guarantees that backward motion mirrors forward motion, preventing policies from favoring a subset of legs when switching direction.

\subsection{Reinforcement Learning and Hardware Setup}
\label{subsec:RLsetup}

We train policies using Proximal Policy Optimization (PPO)~\cite{schulman2017proximal}. 
The actor is a multi-layer perceptron (MLP) with hidden sizes $[512,256,128]$ and ELU activations; 
the critic shares the architecture but outputs a scalar state value. 
Control runs at $50$\,Hz. The hyperparameters of PPO are: learning rate with the initial value $0.001$ and adaptively updated by Adam \cite{adam2014method}, discount $\gamma=0.99$, GAE factor $\lambda=0.95$~\cite{schulman2015high}, clipping threshold $\epsilon=0.2$, and entropy weight $0.01$. 
Training is conducted in Isaac Gym~\cite{makoviychuk2021isaac} with 2048 agents in parallel. 
The policy is updated every 24 simulation steps and training stops after $2.46\times10^9$ samples. 
Domain randomization is applied at initialization and observation noise at each step (Table~\ref{table:domain-randomization}). 
Parameters for the gait-period and duty-factor formulation in Eq.~\ref{eq:gait-period-and-duty-factor-expression} are fixed to $a_T=2.55$, $b_T=0.20$, $c_T=0.975$, $a_\beta=0.5588$, $b_\beta=0.20$, and $c_\beta=0.681$, with the torso height randomized uniformly in $[0.35,0.45]$\,m.

Episodes last $30$\,s with both velocity and gait resampling. 
At the start, forward velocity $v_x^{\text{cmd}}$ is drawn from $[-2,2]$\,m/s, while $v_y^{\text{cmd}}=\omega^{\text{cmd}}_{\text{yaw}}=0$. 
At $t=10$\,s, $v_x^{\text{cmd}}$ is resampled to encourage adaptation. 
For gait selection, an initial gait is chosen from the library with uniform noise $\pm0.02$ on each phase parameter $\theta$; 
at $t=20$\,s, a new gait is sampled to enforce transitions.
Policies are deployed on the Unitree Go2 via wired connection. 
Lightweight Communications and Marshalling~\cite{huang2010lcm} links the robot to a laptop with an RTX~4090 GPU, 
which reads robot and joystick data, executes the trained policy, and sends joint-level commands back. 
No hand-tuned gait schedules are used at deployment; gait commands are provided only through joystick inputs.

\begin{table}[t]
    \centering
    \renewcommand{\arraystretch}{1.2}
    \begin{tabular}{l l l}
        \toprule
        \textbf{Type} & \textbf{Parameter} & \textbf{Range} \\
        \midrule
        \multirow{3}{*}{Dynamics} 
            & Body Mass & $[-1.5,\, 1.5]$ kg \\
            & Body Friction & $[0.3,\, 2.0] \times$ default \\
            & Torso Velocity Perturbation & $[-0.25,0.25]$ m/s \\
        \midrule
        \multirow{3}{*}{State Noises} 
            & Orientation & $[-0.05,\, 0.05]$ \\
            & Joint Position & $[-0.01,\, 0.01]$ rad \\
            & Joint Velocity & $[-1.5,\, 1.5]$ rad/s \\
        \bottomrule
    \end{tabular}
    \caption{Randomized training parameters. Each of the 1024 Isaac Gym environments is assigned randomized dynamics at initialization, and uniform noise is injected into the state vector at every timestep.}
    \label{table:domain-randomization}
    \vspace*{-0.1in}
\end{table}

\subsection{Ablation Study Setup}
We evaluate the contribution of each symmetry component by selectively disabling them during training and testing. To study morphological symmetry, we remove the joint-similarity penalty so that gait evaluation depends only on stance and swing agreement, without enforcing coordination between symmetric limbs. To study time-reversal symmetry, we disable the mapping in Eq.~\ref{eq:time-reversal-map} by omitting the phase flip $\phi$ when $v^{\text{cmd}}_x < 0$, forcing forward and backward motion to share the same phase schedule.

\subsection{Performance Metrics}
Policies are evaluated using the three following metrics:

\subsubsection{Command velocity tracking}
Per-episode root-mean-square normalized error (RMSNE) between commanded and realized base velocities:
\begin{equation}
\mathrm{RMSNE}_{\text{cmd}}
= \left[
\frac{1}{T}\sum_{t=1}^{T}
\sum_{d \in \{x, y, \omega_z\}}
\left(
\frac{v_{d}^{\text{cmd}} - v_{d}(t)}{|v_{d}^{\text{cmd}}|+\varepsilon}
\right)^{2}
\right]^{1/2},
\end{equation}
where $\varepsilon$ is a small constant to avoid division by zero.

\subsubsection{Gait consistency}
Mismatch between desired and realized stance/swing states, with an additional term penalizing asymmetry across symmetric leg pairs.  
The stance consistency term is
\begin{equation}
\mathrm{GC}_{\text{stance}}
= \frac{1}{T}\sum_{t=1}^{T}
\left[
1 - \tfrac{1}{4}\sum_{i=1}^{4}\mathbf{1}\!\big(s_{i}^{\text{real}}(t)=s_{i}^{\text{des}}(t)\big)
\right],
\end{equation}
and the morphological consistency term is
\begin{equation}
\mathrm{GC}_{\text{morph}}
= \frac{1}{T}\sum_{t=1}^{T} \sum_{(a,b)\in\mathcal{P}} \|q_a(t) - q_b(t)\|,
\end{equation}
where $\mathcal{P}$ denotes selected symmetric leg pairs and $q_a$ the joint positions.  
The combined measure is
\begin{equation}
\mathrm{GC}
= \frac{1}{T}\sum_{t=1}^{T}\big(\mathrm{GC}_{\text{stance}}(t)+\mathrm{GC}_{\text{morph}}(t)\big).
\end{equation}

\subsubsection{Cost of transport (CoT)}
Normalized positive mechanical work relative to weight and distance traveled:
\begin{equation}
\mathrm{CoT}
= \frac{\int_{0}^{T}\sum_{j}\big|\tau_{j}(t)\,\dot{q}_{j}(t)\big|_{+}\,dt}
{m g \,\Delta x},
\end{equation}
where $\tau_j$ and $\dot{q}_j$ denote the torque and joint velocity of actuator $j$, $mg$ is the robot’s weight, and $\Delta x$ is the net forward displacement over the episode.

\section{Results}
\label{results}

We assess the proposed symmetry-guided RL framework in simulation and hardware. The evaluation covers: (A) velocity tracking across commanded speeds, (B) gait tracking and transitions among trotting, bounding, half-bounding, and galloping, (C) ablation of morphological symmetry, and (D) hardware validation on the Unitree Go2. These results demonstrate accurate tracking, versatile gait generation, and reliable sim-to-real transfer.

\begin{figure}[ht]
\centering
\includegraphics[width=0.95\columnwidth]{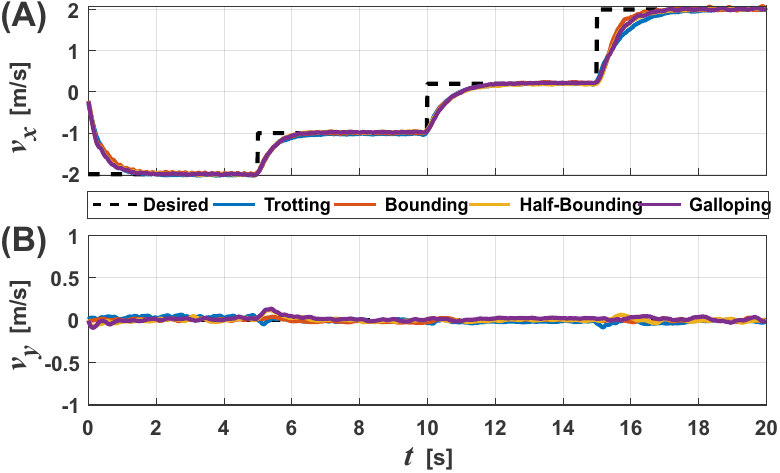}
\caption{
Velocity tracking across four gaits under varying commanded forward speeds. 
(A) Sequential transitions with $v_{x}^{\text{cmd}} \in \{-2,\,-1,\,0.2,\,2\}$~[m/s]. 
(B) Tracking of $v_{y}^{\text{cmd}}=0$ during the same test.}
\label{fig:tracking_vel_adj}
\vspace*{-0.15in}
\end{figure}

\begin{figure*}[t]
\centering
\includegraphics[width=1.0\textwidth]{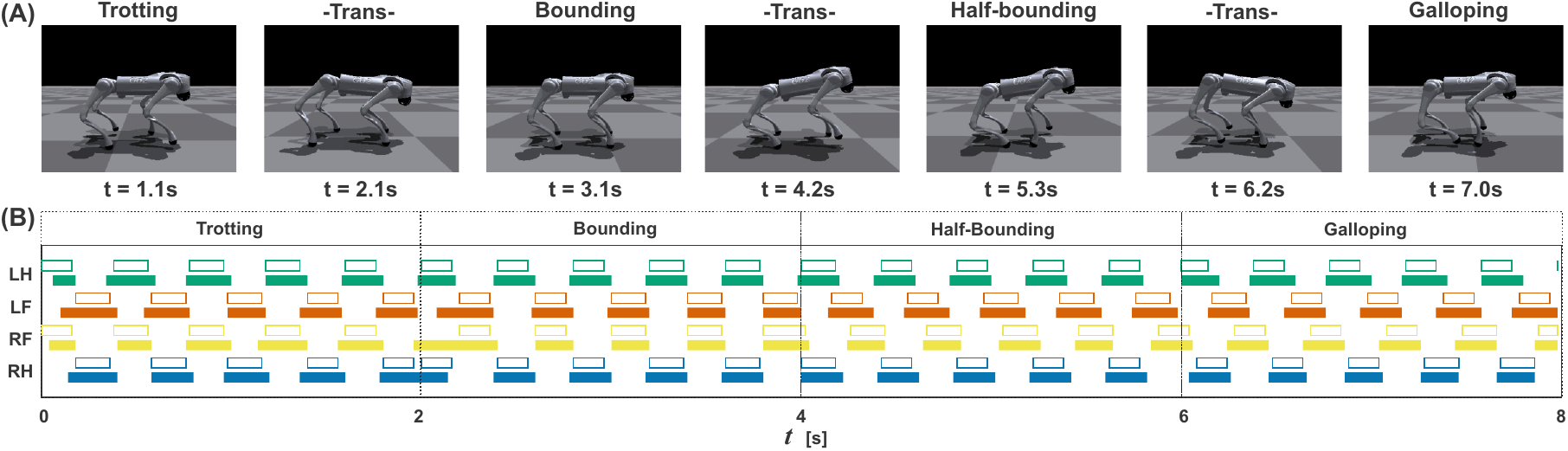}
\caption{
Gait tracking at $v_{x}^{\text{cmd}}=0.5~[\text{m/s}]$.  
(A) Representative frames of trotting, bounding, half-bounding, and galloping, including transition phases.  
(B) Desired footfall sequences (hollow bars) compared with realized touchdown sequences in simulation (solid bars).}
\label{fig:track_gait}
\vspace*{-0.2in}
\end{figure*}

\subsection{Velocity Tracking}
We first evaluated velocity tracking in simulation across four gaits. 
The commanded forward velocity $v_{x}^{\text{cmd}}$ was initialized at $-2~[\text{m/s}]$ and changed sequentially to $-1$, $0.2$, and $2~[\text{m/s}]$ at $t=5$, $10$, and $15~[\text{s}]$, respectively, while the lateral command was fixed at $v_{y}^{\text{cmd}}=0$. 
As shown in Fig.~\ref{fig:tracking_vel_adj}, the actual velocity converged to each commanded value within approximately $1~[\text{s}]$, demonstrating stable and responsive tracking. 
Performance was consistent across all gaits, indicating that the learned policy achieves equally accurate velocity regulation independent of footfall sequence.

\subsection{Gait Tracking}
We evaluated gait tracking under a constant forward velocity of $v_{x}^{\text{cmd}}=0.5~[\text{m/s}]$. 
The robot sequentially executed trotting, bounding, half-bounding, and galloping, each lasting $2~[\text{s}]$. 
Fig.~\ref{fig:track_gait}(A) shows representative frames of the four gaits, including transition phases. 
Fig.~\ref{fig:track_gait}(B) compares desired footfall sequences (hollow) with simulated contact sequences (solid). 
All transitions were achieved within a single step (approximately $0.5~[\text{s}]$). 
The mean absolute error between desired and realized contacts was $\text{GC}_{\text{stance}}=0.07$, indicating accurate tracking. 
The learned policy generalized across arbitrary gait transitions and commanded velocities, while maintaining stable body posture without collapse or lateral drift. 
Supplementary demonstrations, including left- versus right-leading leg switches in half-bounding and galloping, and abrupt transitions from galloping to trotting, are provided in the online repository\footnote{Supplementary material and videos: \url{https://anonymous.4open.science/r/go2_symm_rl-F623/}}.

\subsection{Ablation Study Results}
\label{subsec:ablation_results}
\begin{figure}[h]
\centering
\includegraphics[width=0.95\columnwidth]{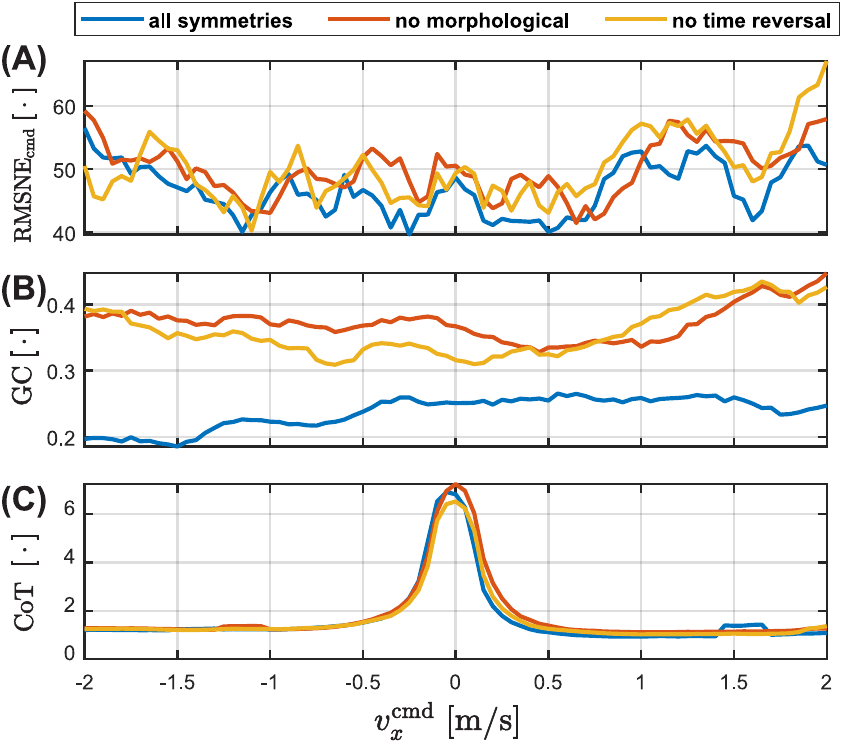}
\caption{
Effect of ablating morphological/time-reversal symmetry: (A) command velocity tracking, (B) gait consistency, and (C) CoT. The symmetry-enforced policy consistently outperforms the no-symmetry variant in accuracy, coordination, and efficiency.}
\label{fig:ablation}
\vspace*{-0.2in}
\end{figure}

We assessed the role of morphological and time-reversal symmetries by comparing the full symmetry-enforced policy (ours) against variants trained with individual symmetry terms removed. Performance was evaluated using three metrics averaged across all gaits introduced in Sec.~\ref{subsec:gait_spec}, as summarized in Fig.~\ref{fig:ablation}. 

\textit{Command velocity tracking}: The full policy achieved the lowest RMSNE across positive speeds, demonstrating the most accurate forward tracking. Removing morphological symmetry caused noticeable degradation, while removing time-reversal symmetry mainly increased errors at negative speeds, confirming its importance for backward motion.  

\textit{Gait consistency}: The full policy achieved an average $\text{GC}$ error of about $0.2$, whereas the no-morphological variant rose to about $0.4$. Removing time-reversal symmetry had little impact on forward gaits but reduced consistency during reversals. These results highlight that morphological symmetry primarily governs inter-leg coordination, while time-reversal symmetry ensures mirrored behaviors when switching direction.

\textit{Cost of transport}: The full policy achieved comparable 
CoT to both the no-morphology and no-time-reversal variants when moving at negative velocities, but showed consistently lower CoT for positive velocities. For instance, at $v_{\text{cmd}} = 0.2\ \text{m/s}$, the full policy reached $\text{CoT} = 2.86$, compared to $3.47$ for the no-time-reversal and $4.15$ for the no-morphology variants. At $v_{\text{cmd}} = 0.5\ \text{m/s}$, the three policies yielded CoT values of $1.10$, $1.25$, and $1.39$, respectively. Overall, the no-time-reversal variant incurred roughly $10$--$20\%$ higher energy cost than the full policy, while the no-morphology variant was $30$--$50\%$ more costly. These results indicate that morphological symmetry has the strongest influence on energetic efficiency, whereas time-reversal symmetry primarily enhances directional robustness.

Overall, these results show that both morphological and time-reversal symmetries play distinct but complementary roles. Morphological symmetry drives coordination and efficiency, while time-reversal symmetry guarantees consistent bidirectional locomotion. Their combination yields the most accurate, coordinated, and efficient gaits.

\subsection{Hardware Performance}
\begin{figure*}[t]
\centering
\includegraphics[width=1.0\textwidth]{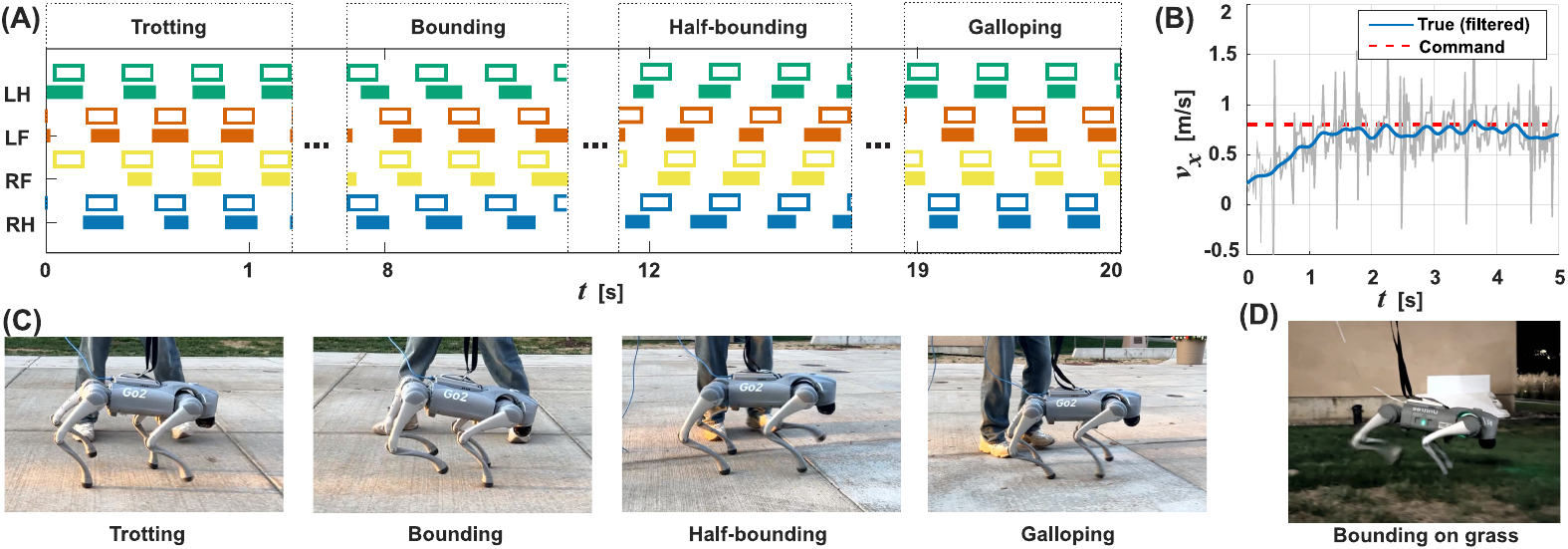}
\caption{
Hardware validation on the Unitree Go2:
(A) Comparison of desired (hollow) and true (solid) contact sequences across trotting, bounding, half-bounding, and galloping. (B) Velocity tracking at $v_{x}^{\text{cmd}} = 0.8~[\text{m/s}]$, where the true velocities are predicted by a state estimator. (C) Keyframes of robot applying the four gaits. (D) Robustness test on uneven terrain, where bounding gaits remained stable during transitions from concrete to grass.}
\label{fig:hardware_performance}
\vspace*{-0.2in}
\end{figure*}

We validated the learned policy on the Unitree Go2. A constant forward command of $v_{x}^{\text{cmd}} = 0.8~[\text{m/s}]$ was applied, while gaits were switched sequentially from trotting to bounding, half-bounding, and galloping, each lasting $5~[\text{s}]$.  
Figure~\ref{fig:hardware_performance}(A) compares desired and realized contact sequences. In all cases the robot reproduced the expected footfall patterns. For trotting and galloping, contact timing closely matched the reference. In bounding and half-bounding, the paired limbs maintained correct synchronization and phasing, but contacts deviated by about $0.1~[\text{s}]$ from the desired sequence. Despite these shifts, all gaits executed stably without collapse or lateral drift.  
Keyframes of the four gaits are shown in Fig.~\ref{fig:hardware_performance}(B) consistent with the desired sequences. The robot converged to the commanded velocity within $1~[\text{s}]$, demonstrating reliable tracking. Robustness was further tested by executing bounding on uneven terrain: as shown in Fig.~\ref{fig:hardware_performance}(D), the robot transitioned smoothly from concrete to grass without failure, confirming that the policy generalizes to outdoor conditions and maintains stability under perturbations.

\section{Conclusions}
This work introduced a symmetry-guided reinforcement learning framework that unifies quadrupedal gait generation without relying on predefined trajectories or gait-specific controllers. By embedding temporal, morphological, and time-reversal symmetries directly into the reward design, a single policy reproduced trotting, bounding, half-bounding, and galloping on the Unitree Go2 robot, achieving accurate velocity tracking, coordinated footfall patterns, and improved energetic efficiency. The use of period and duty factor sampling curves further enabled scalable gait modulation, supporting more dynamic locomotion across speeds. Simulation and hardware results confirmed that symmetries enhance the adaptability and robustness across diverse gaits.

\bibliographystyle{IEEEtran}
\bibliography{Ref_Jiayu_v2}

\end{document}